\documentclass[prl,twocolumn,preprintnumbers,amsmath,amssymb]{revtex4}

\usepackage{graphicx}% Include figure files
\usepackage{bm}% bold math
\usepackage{color}% bold math

\begin{document}

%\preprint{APS/123-QED}

\title{Two-dimensional square lattice polonium stabilized by the spin-orbit coupling}
\author{Shota Ono}
%\email{shota_o@gifu-u.ac.jp} % revtex
\email{shota\_o@gifu-u.ac.jp} % arXiv
\affiliation{Department of Electrical, Electronic and Computer Engineering, Gifu University, Gifu 501-1193, Japan}

\begin{abstract}
Polonium is known as the only simple metal that has the simple cubic (SC) lattice in three dimension. There is a debate about whether the stabilized SC structure is attributed to the scalar relativistic effect or the spin-orbit coupling (SOC). Here, we study another phase, two-dimensional (2D) polonium (poloniumene), by performing density-functional theory calculations. We show that the 2D polonium has the square lattice structure as its ground state and demonstrate that the SOC (beyond the scalar relativistic approximation) suppresses the Peierls instability and is necessary to obtain no imaginary phonon frequencies over the Brillouin zone. 
\end{abstract}

%\pacs{61.48.-c, 72.80.Rj, 73.20.Fz}
%61.48.-c: Structure of fullerenes and related hollow and planar molecular structures
%72.80.Rj: Fullerenes and related materials / Conductivity of specific materials
%73.20.Fz: Weak or Anderson localization / Electron states at surfaces and interfaces

\maketitle

%%%%%%%%%%%%%%%%%%%%%%%%%%%%%%%%%%%%
{\it Introduction.---}Since the discovery of polonium (Po) by Marie and Pierre Curie in 1898, many researches have been made about the physical and chemical properties. The intriguing property of Po is that it has the simple cubic (SC) lattice, $\alpha$-Po phase, as its ground state, quite different from the fact that most elementary metals have face-centered cubic, body-centered cubic, and hexagonal closed packed lattices. Since Po shows a strong radioactivity and is therefore dangerous to humans, theoretical and computational studies are of importance in understanding the origin of the stabilized SC phase. The density-functional theory (DFT) studies have revealed that the scalar relativistic or spin-orbit coupling (SOC) terms are responsible for the stabilization of the SC phase \cite{min,legut,kim,sob,verstraete,belabbes,kang}. For example, Legut {\it et al}. have shown that the scalar relativistic effect is enough for SC Po to be stabilized \cite{legut}, whereas Min {\it et al}. have shown that the SOC is more important to prevent the phonon softening from taking place \cite{min,kang}. In this paper, we investigate the two-dimensional (2D) case in order to study how the SOC affects on the lattice stability. 

The square (SQ) lattice in two-dimension is an analog of the SC lattice in three-dimension. Recently, Nevalaita and Koskinen have performed systematic DFT calculations of 2D metals up to the atomic number of $Z=83$, i.e., bismuth \cite{nevalaita}. However, they have shown that most metals are energetically stable in the hexagonal and honeycomb lattices, but unstable in the SQ lattice geometry. An exception is the SQ lattice of ruthenium, which is more stable than the hexagonal and honeycomb lattices by 0.04 eV and 0.38 eV per atom, respectively, while the former energy is comparable to the thermal energy. 

In this paper, we study the 2D Po, {\it poloniumene}, by performing DFT calculations and demonstrate that the poloniumene has the SQ lattice as its ground state. We show that the SOC plays a crucial role in obtaining the phonon band structure with no imaginary phonon frequencies over the Brillouin zone (BZ). By calculating the susceptibility and the 2D Fermi surface with and without the SOC, we show that the SOC suppresses the Peierls instability, giving a dynamical stability to the SQ lattice of poloniumene. 

It has been known that the SOC is important for an accurate description of the phonon band structure of heavy metals, such as three-dimensional (3D) \cite{sanchez} and 2D bismuth (bismuthene) \cite{akturk}. It is interesting to note that the dynamical stability of metastable platinum with hexagonal closed packed lattice is endowed with SOC \cite{schonecker}. In the present work, we also emphasize that the SOC is mandatory for obtaining no imaginary frequencies in poloniumene. The present study will improve the fundamental understanding of the SOC effect on the lattice dynamics of heavy metals.

The SQ lattice structure is quite rare in two-dimension. Zhao {\it et al}. have reported a synthesis of 2D iron having a SQ lattice in graphene pores \cite{zhao}. However, Shao {\it et al}. have shown that such a structure is unstable by using DFT \cite{shao}. Recently, Kano {\it et al}. have reported that on graphene 2D copper oxide has a SQ lattice \cite{kano}, while this is stabilized by the presence of the oxygen atom located at the center of the unit cell. The poloniumene proposed in the present work serves as a valuable example with a rare structure and can have a strong impact in the field of 2D materials.  

%%%%%%%%%%%%%%%%%
\begin{table*}
\begin{center}
\caption{The minimum value of the total energy $E_{\rm tot}$ (eV/atom) at the equilibrium interatomic distance $a_{0}$ (\AA) for SQ, HX, and HC structures and for the cases without and with the SOI. }
{
%\begin{tabular}{lcccccc}\hline
\begin{tabular}{|l|c|c|c|c|c|c|}\hline
%--------------------------------------------------------------------------------
   & SQ (woSOC) & HX (woSOC)  & HC (woSOC) & SQ (wSOC) & HX (wSOC) & HC (wSOC) \\ \hline
$E_{\rm tot}$  & 0.0 & 0.234  & 0.240  & 0.0  & 0.142  & 0.210  \\ \hline
$a_0$ & 3.175  & 3.338  & 3.107  & 3.258 & 3.460  & 3.187  \\ \hline
%--------------------------------------------------------------------------------
\end{tabular}
}
%--------------------------------------------------------------------------------
%{
%\begin{tabular}{lcccccc}\hline
%   & SQ (woSOI) & HX (woSOI)  & HC (woSOI) \hspace{5mm} & SQ \hspace{5mm}  & HX \hspace{5mm}   & HC \hspace{5mm}   \\ \hline
%$E_{\rm tot}$ \hspace{5mm}  & 0.0 & 0.234  & 0.240 \hspace{5mm} & 0.0 \hspace{5mm} & 0.142 \hspace{5mm} & 0.210 \hspace{5mm}  \\ \hline
%$a_0$ \hspace{5mm} & 3.175  & 3.338  & 3.107 \hspace{5mm}  & 3.258 \hspace{5mm} & 3.460 \hspace{5mm} & 3.187 \hspace{5mm}  \\ \hline
%\end{tabular}
%}
%--------------------------------------------------------------------------------
\label{table1}
\end{center}
\end{table*}
%%%%%%%%%%%%%%%%%

%%%%%%%%%%%%%%%%%%%%%%%%%%%%%%%%%%%
{\it Computational details.---}To perform DFT calculations, we use the plane-wave based program of Quantum ESPRESSO \cite{qe}. The effects of exchange and correlation are treated within both LDA \cite{pz} and GGA \cite{pbe}. The core electrons are treated within the PAW method \cite{kresse}. The valence configuration of Po is $(5d)^{10}(6s)^2(6p)^4$. The cutoff energies for the wavefunction and the charge density are 100 (80) Ry and 1000 (800) Ry with (without) SOC, respectively. The self-consistent calculations are performed by using 30$\times$30$\times$1 $k$ Monkhorst-Pack grids \cite{MK} and the smearing parameter of 0.025 Ry. The vacuum region between the layers is taken to be larger than 15 \AA. The geometry optimization is performed for the phonon calculations, where the total energy is converged within $10^{-5}$ Ry. The phonon band structure calculations are performed by using 4$\times$4$\times$1, 6$\times$6$\times$1, and 8$\times$8$\times$1 $q$ grids. We have confirmed that at least 8$\times$8$\times$1 $q$ grids are necessary to obtain no imaginary frequencies over the BZ for the calculations with SOC. 

To investigate the Peierls instability of the poloniumene, we calculate the noninteracting susceptibility at the wavevector $\bm{q}$
\begin{eqnarray}
 \chi(\bm{q}) = - \frac{1}{N_{\rm c}}\sum_{n,n'}\sum_{\bm{k}} 
 \frac{f(\varepsilon_{n'\bm{k}+\bm{q}}) - f(\varepsilon_{n\bm{k}}) }
 {\varepsilon_{n'\bm{k}+\bm{q}} - \varepsilon_{n\bm{k}}},
 \label{eq:response}
\end{eqnarray}
where the matrix elements are neglected \cite{gruner}. $\varepsilon_{n\bm{k}}$ is the electron energy for the wavevector $\bm{k}$ and the band index $n$. For the case without SOC, the spin index is included to $n$. $f(\varepsilon)$ is the Fermi distribution function at zero temperature for the electron energy $\varepsilon$. $N_{\rm c}$ is the number of unit cell. For the summation with respect to $\bm{k}$ in Eq.~(\ref{eq:response}), a 100$\times$100$\times$1 $k$ mesh is used. 

Below we demonstrate how the SQ lattice is stabilized in poloniumene by calculating the total energy, the phonon band structure, and the noninteracting susceptibility as well as the 2D Fermi surface. We show the DFT results based on the generalized gradient approximation (GGA) only, while the same scenario also holds (the SQ lattice is stabilized by SOC) when the local-density approximation (LDA) is used. 

%%%%%%%%%%%%%%%%%
\begin{figure}[ttt]
\center
\includegraphics[scale=0.35]{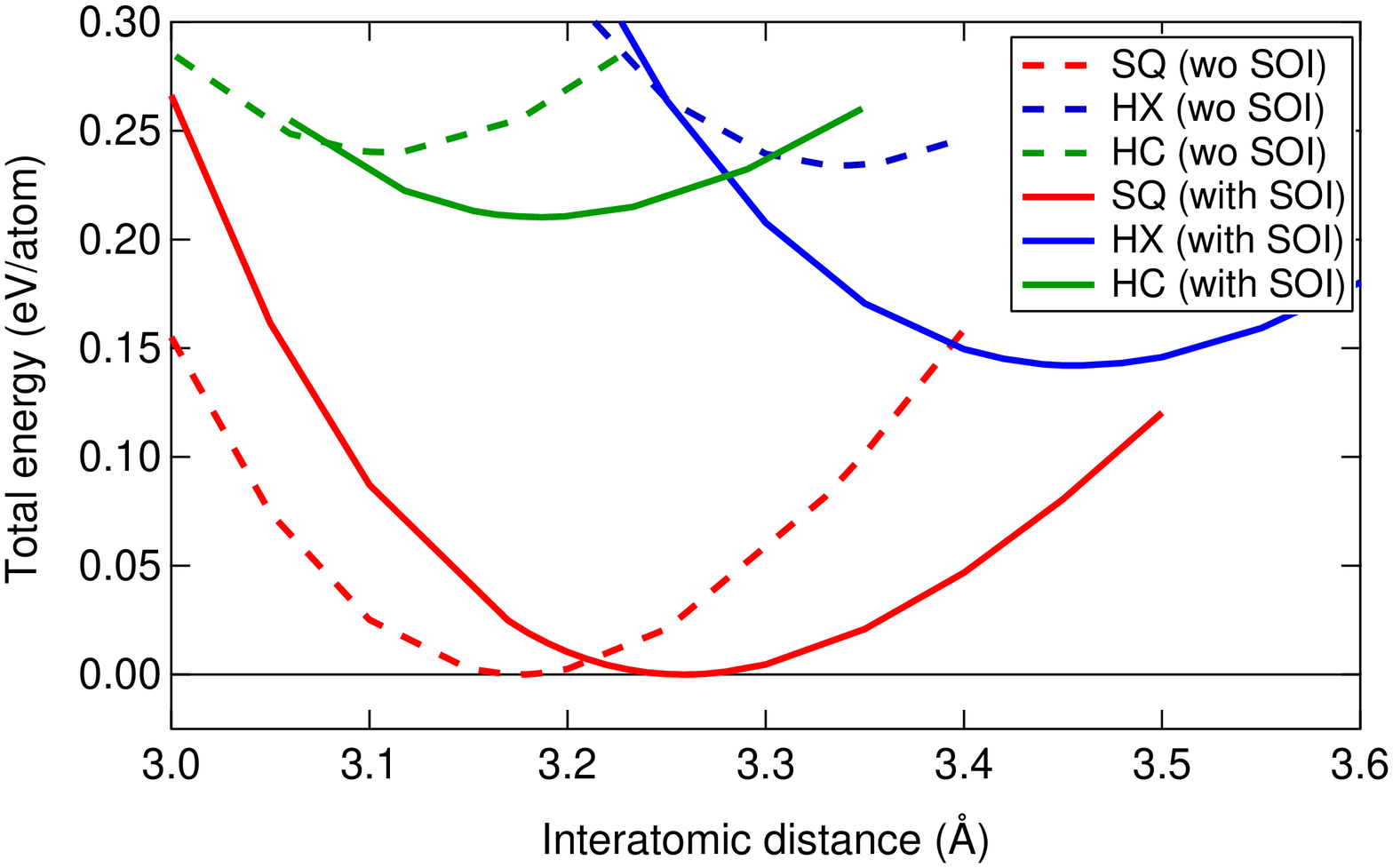}
\caption{\label{fig1} The total energy versus the interatomic distance for SQ, HX, and HC structures of 2D Po. The total energy is measured from the minimum energy of the SQ structure. }
\end{figure}
%%%%%%%%%%%%%%%%%

%%%%%%%%%%%%%%%%%
\begin{figure}[ttt]
\center
\includegraphics[scale=0.4]{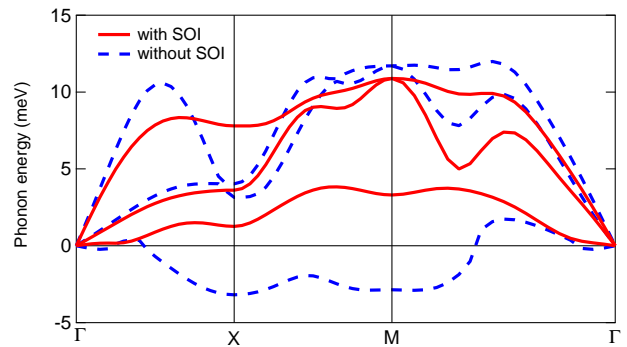}
\caption{\label{fig2} The phonon band structure of poloniumene having the SQ lattice: with SOC (solid) and without SOC (dashed). The imaginary phonon frequency is expressed by negative phonon energy.}
\end{figure}
%%%%%%%%%%%%%%%%%

%%%%%%%%%%%%%%%%%
\begin{figure}[ttt]
\center
\includegraphics[scale=0.35]{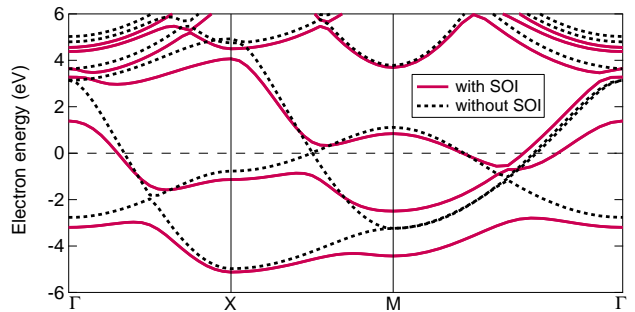}
\caption{\label{fig3} The electron band structure of the SQ lattice of poloniumene, with and without the SOC in DFT calculations. The electron energy is measured from the Fermi level.}
\end{figure}
%%%%%%%%%%%%%%%%%

%%%%%%%%%%%%%%%%%
%\begin{figure}[ttt]
%\center
%\includegraphics[scale=0.35]{Fig_4.eps}
%\caption{\label{fig4} The projected DOS of 2D Po around the Fermi level. }
%\end{figure}
%%%%%%%%%%%%%%%%%

%%%%%%%%%%%%%%%%%
\begin{figure}[ttt]
\center
\includegraphics[scale=0.35]{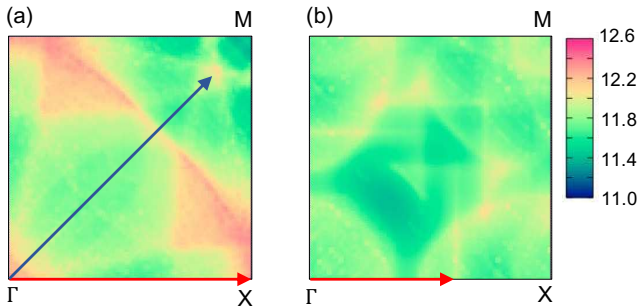}
\caption{\label{fig4} $\chi (\bm{q})$ (states/eV) for poloniumene: (a) without and (b) with the SOC. Examples of the nesting vectors are indicated by arrows: $\bm{q}={\rm X}$ and $0.87$M for (a) and $\bm{q}=0.6{\rm X}$ for (b). }
\end{figure}
%%%%%%%%%%%%%%%%%

%%%%%%%%%%%%%%%%%
\begin{figure}[ttt]
\center
\includegraphics[scale=0.35]{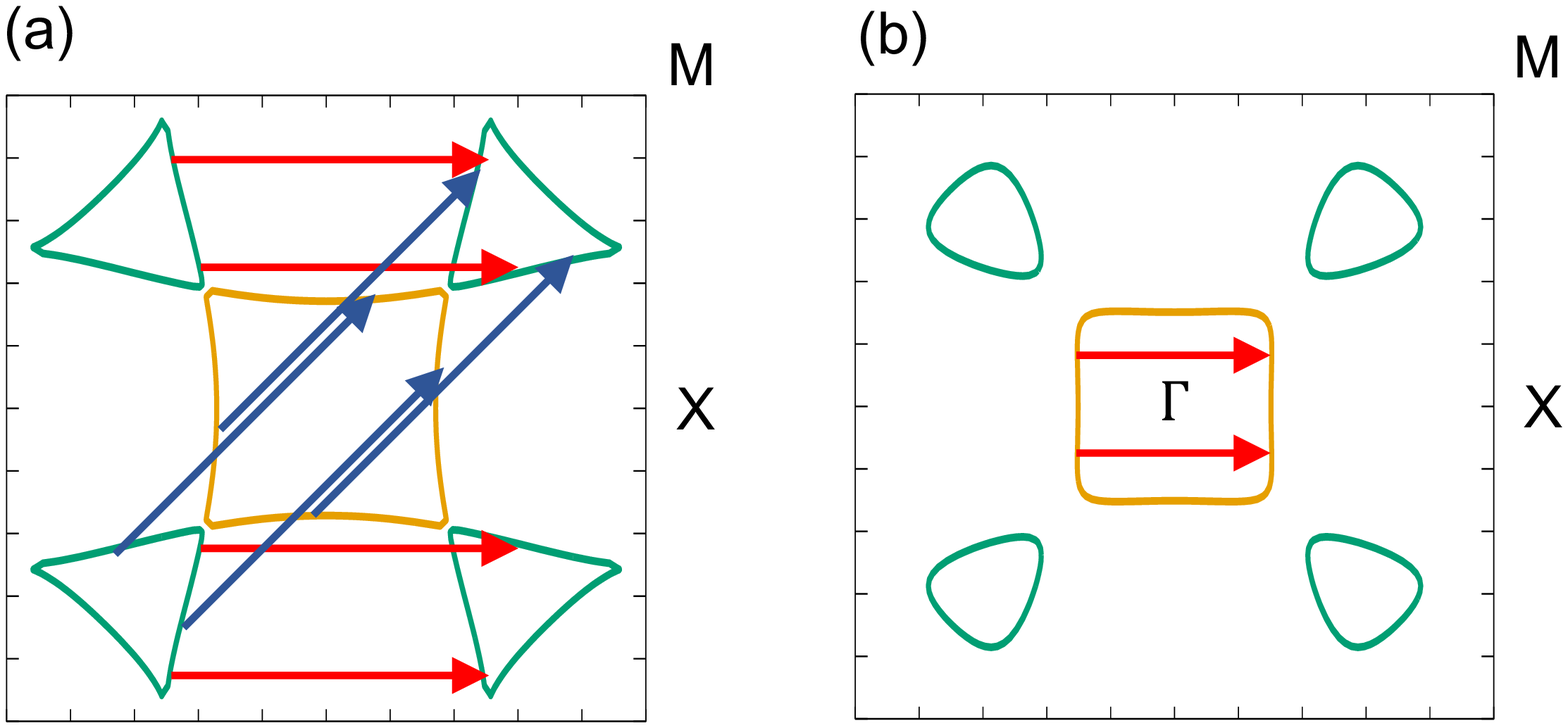}
\caption{\label{fig5} The Fermi line for poloniumene: (a) without and (b) with the SOC. The center corresponds to the point $\Gamma$. The nesting vectors indicated correspond to those in Fig.~\ref{fig4}. }
\end{figure}
%%%%%%%%%%%%%%%%%

%%%%%%%%%%%%%%%%%
\begin{figure}[ttt]
\center
\includegraphics[scale=0.3]{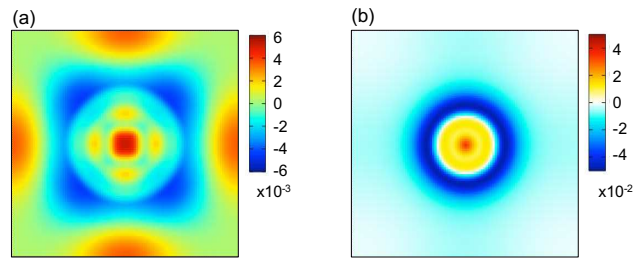}
\caption{\label{fig6} The total charge density (a.u.) with the SOC in a unit cell, subtracted by (a) the superposition of atomic density and (b) the total charge density without the SOC.}
\end{figure}
%%%%%%%%%%%%%%%%%

%%%%%%%%%%%%%%%%%%%%%%%%%%%%%%%%%%%
%\section{Results and Discussion}
{\it Stable structure.---}Figure \ref{fig1} shows the total energy (per atom) as a function of the interatomic bond length for three lattice structures: the SQ, hexagonal (HX), and honeycomb (HC) lattices. The curves are calculated without and with the SOC. For both cases, the total energy is measured from that of the most stable geometry of the SQ lattice. The cohesive energy (without SOC) of the SQ lattice is estimated to be 2.58 eV/atom, which is smaller than that of the SC lattice (2.73 eV/atom). The SQ structure is more stable than the HX and HC structures by 0.142 (0.234) eV and 0.210 (0.240) eV with (without) SOC, respectively. These values are larger than the thermal energy of room temperature by an order of magnitude, implying no phase coexistence at ambient conditions. The minimum value of total energy $E_{\rm tot}$ and the corresponding interatomic distance $a_0$ are listed in Table \ref{table1}. The size of $a_0$ increases slightly when the SOC is included, similar to Ref.~\cite{legut}. As the coordination number decreases (HX$\rightarrow$SQ$\rightarrow$HC), the value of $a_0$ also decreases, so that the lattice constant of the SQ lattice is shorter than that of the SC lattice of 3.359 \AA \ \cite{desando}. 

{\it Phonons.---}In order to study the dynamical stability of the SQ lattice, we perform the phonon band structure calculations with and without the SOC. Figure \ref{fig2} shows the dispersion curves along the symmetry lines in the BZ, where the imaginary phonon energy is expressed by negative values. Without the SOC, the imaginary frequencies appear around high symmetry point, X and M, indicating that the SQ lattice structure is unstable. With the SOC, on the other hand, no imaginary frequencies appear, clearly indicating that the SOC effect is important to stabilize the SQ structure of poloniumene. This scenario is similar to the case of 3D Po, where the SOC plays a key role for the realization of the SC lattice \cite{kang}. 

We note some characteristic properties of phonons in poloniumene. Around $\Gamma$, there are three acoustic phonons: the longitudinal acoustic, transverse acoustic, and flexural modes. The velocity of the former two branches with a linear dispersion, $v_{\rm L}$ and $v_{\rm T}$, is obtained as follows: $v_{\rm L}^{\rm X}=3.4$ km/s and $v_{\rm T}^{\rm X}=0.9$ km/s along $\Gamma$-X direction and $v_{\rm L}^{\rm M}=2.8$ km/s and $v_{\rm T}^{\rm M}=2.1$ km/s along $\Gamma$-M direction. These are related to the elastic constants ($c_{11}$, $c_{12}$, and $c_{44}$) as follows: $v_{\rm L}^{\rm X}=\sqrt{c_{11}/\rho}$, $v_{\rm T}^{\rm X}=\sqrt{c_{44}/\rho}$, $v_{\rm L}^{\rm M}=\sqrt{(c_{11}+c_{12}+ 2c_{44})/\rho}$, and $v_{\rm T}^{\rm M}=\sqrt{(c_{11}-c_{12})/\rho}$ with $\rho$ the mass density per area \cite{kittel}. Assuming $\rho = 3.28\times 10^{-6}$ kg/m$^2$, one obtains $c_{11}= 38.2$, $c_{12}\simeq 8.4$, and $c_{44}= 2.7$ GPa nm. The elastic anisotropy of the poloniumene is quite large (i.e., $c_{11}\gg c_{44}$), compared to other 2D metals \cite{nevalaita}, but is consistent with the case of 3D Po having the SC lattice \cite{legut}. 

The origin of the anisotropy of the elastic constants obtained is an intrinsic property of the SQ lattice. To show it qualitatively, we consider a central potential, $V(R)$, with $R$ the interatomic distance up to the second nearest-neighbour (NN) atoms. By diagonalizing the dynamical matrix \cite{AM}, one obtains $v_{\rm L}^{\rm X} = a_{0} \sqrt{(\xi_1 + \xi_2 + \eta_2)/M}$ and $v_{\rm T}^{\rm X} = a_{0} \sqrt{(\xi_2 + \eta_2)/M}$, where $M$ is the mass of Po atom, $\xi_l = V''(R_l)$, and $\eta_l = V'(R_l)/R_l$ for $l=1,2$. The derivative (prime) of $V$ with respect to $R$ is evaluated at $R_l$ the distance of the $l$th NN atoms, i.e., $R_1=a_0$ and $R_2=\sqrt{2}a_0$. Notice that the expression of $v_{\rm T}^{\rm X}$ vanishes when $\xi_2$ and $\eta_2$ are neglected. Since the contribution from $l=2$ is generally smaller than that from $l=1$, the inequality $v_{\rm L}^{\rm X} \gg v_{\rm T}^{\rm X}$ is obtained. A similar discussion holds for the phonon energy at the point N for body-centered cubic lattice \cite{ono2019}. 

{\it Electrons.---}Figure \ref{fig3} shows the electron band structure of poloniumene having the SQ lattice, for the cases with (solid) and without (dashed) the SOC. For the latter, the band crossings occur at the middle of the lines of $\Gamma$-X, X-M, and M-$\Gamma$. Such bands split when the SOC is added, and hence the Fermi level crosses the dispersive band along the lines of $\Gamma$-X and M-$\Gamma$ only. These bands depicted are formed from $6p$-states mainly, while the $6s$ band is located well below the Fermi level by more than 10 eV and decoupled from the $6p$ bands. 

{\it Peierls instability.---}To understand the origin of the dynamical stability of poloniumene, we calculate $\chi(\bm{q})$ in Eq.~(\ref{eq:response}) using $\varepsilon_{n\bm{k}}$ obtained from DFT calculations. Figures \ref{fig4}(a) and \ref{fig4}(b) show $\chi(\bm{q})$ for poloniumene without and with the SOC, respectively. Without the SOC, the value of $\chi(\bm{q})$ is strongly enhanced around $\bm{q}={\rm X}$ and $0.87$M, at which in Fig.~\ref{fig2} the phonon softening and imaginary frequencies are observed. With the SOC, on the other hand, such a peak in $\chi(\bm{q})$ is smeared out, so that positive phonon energies are obtained in Fig.~\ref{fig2}. 

The $\bm{q}$-dependence of $\chi$ can be understood in terms of the Fermi surface nesting in two-dimension. Figures \ref{fig5}(a) and \ref{fig5}(b) show the Fermi lines without and with the SOC, respectively. The almost square hole pocket around $\Gamma$ as well as the electron pockets located along the $\Gamma$-M direction are shrinked when the SOC is included. Accordingly, it is difficult to find the nesting vectors on the Fermi lines. In this way, the SOC suppresses the Peierls instability in poloniumene having the SQ lattice. 

It would be helpful to discuss the Peierls instability in view of the charge density analysis. Figure \ref{fig6}(a) shows the contour plot of the charge density minus the superposition of atomic densities with the SOC, where the Po atom is located at the center of the unit cell. The bonding between the NN sites is strongly enhanced and the SQ lattice symmetry is clearly observed because four $6p$ electrons are occupied below the Fermi level. Figure \ref{fig6}(b) shows the charge density with the SOC minus that without the SOC. This shows that the effect of the SOC enhances the electron density at the nuclei by $0.033$ a.u., but lowers that at the middle of the NN sites by $0.005$ a.u. This weakens the directional bonding between Po atoms and elongates the interatomic bond length, as listed in Table \ref{table1}, while this prevents the poloniumene from the Peierls instability. Similar scenario holds for explaining the origin of the stabilized SC structure of Po in three dimension \cite{min,kang}. 

It has been known that the Peierls instability is prone to occur in low dimensional systems \cite{gruner}. This is because the value of $\chi (\bm{q})$ is easily enhanced for low dimension when the nesting vectors are present at the Fermi surface. The Peierls instability has been addressed to occur in 3D Po \cite{min,verstraete,kang} because many phonon soft modes are observed in the DFT calculations without SOC, whereas such modes are hardened with SOC included. The debate on the origin of the stabilized SC phase in Po may be replaced with a question how the Peierls instability is suppressed enough to yield no imaginary phonon frequencies. We have demonstrated that in 2D Po the Peierls instability is suppressed not by the scalar relativistic terms but by SOC. 

%%%%%%%%%%%%%%%%%%%%%%%%%%%%%%%%%%%
%{\it Summary.---}As a 3D counterpart, we have investigated the properties of 2D Po, poloniumene, having the SQ lattice by using DFT. In the phonon band structure calculations, we have shown that the SOI is necessary to obtain the dynamical stability of the poloniumene. This is because the SOI suppresses the Peierls instability of the SQ lattice. Our finding that the SOI is responsible for the stability in 2D as well as 3D Po \cite{kang} is quite important for a deep understanding of the fundamental properties in heavy metals.

%Most 2D materials with SQ-like lattices can be created on an appropriate substrate \cite{kano} or in a nanopore \cite{zhao,shao}. Since the 3D Po can be created by the neutron irradiation of bismuth, the poloniumene may be produced from bismuth layers \cite{nagao,kawakami}.

\begin{acknowledgments}
This study is supported by the Nikki-Saneyoshi Foundation. 
\end{acknowledgments}

\end{document}